\documentstyle[ptpbs1]{ptptex}          

\title
{Study of  effective interaction from single particle transfer reactions
on f-p shell nuclei}
\author{
{H. Sharda\thanks{shardah@usa.net}, R. K. Bansal and Ashwani Kumar$^{\ast}$ }\\
{\it Department of Physics, Panjab University,}\\
{\it Chandigarh -160014, India}\\
{\it $^\ast$Department of Applied Sciences, Punjab Engineering College,}\\
{\it Chandigarh - 160012, India} }

\begin{document}
\maketitle
\begin{abstract}
\baselineskip 24pt
The present study concentrates on the average effective two-body
interaction matrix elements being extracted,  using sum-rule techniques,
from transfer reactions on target states having  single orbital as well as
two orbital 
occupancy. This investigation deals with transfer  reactions on f-p
shell nuclei involving (i) $1f_{7/2}$ and $2p_{3/2}$ transfer  on target
states using $^{40}$Ca as inert core, and (ii) $2p_{3/2}$ and$1f_{5/2}$
transfer on states using $^{56}$Ni as core.
\end{abstract}
\section {Introduction}
The conventional  search for an effective two-body interaction to be
used in shell-model studies involves numerically intensive mixed
configuration calculations.\\
It is, however, now known that meaningful averages of effective
interaction matrix elements can be obtained through an alternative
simpler method with the help of energy-weighted sum
rules\cite{r1,r2,r3,r4,r5,r6}.
Monopole
energy weighted  sum rules have been used to derive explicit algebraic
equations relating effective interaction matrix  elements to isospin
centroids of residual nuclear states obtained via single nucleon
stripping and pick-up reactions performed on general multishell target
states\cite{r4,r5}. With the help of these, we have previously calculated the
$1d^2_{5/2}$, $2s^2_{1/2}$, $1d^2_{3/2}$ and $1f^2_{7/2}$ matrix
elements of average effective interaction, working in each case within a
limited vector space comprising  of just one active orbit outside the
postulated inert core\cite{r5,r6}.\\
The present work reports the application of these sum rules to
stripping and pick-up reactions on target states belonging to f-p shell
nuclei with shell occupancy no longer restricted only to single active orbit
but also extended to two active
orbits. The  target states  included in the current study have either
$1f_{7/2}$ orbit or both $1f_{7/2}$ and $2p_{3/2}$  active orbits
outside the inert core, $^{40}$Ca. We also study the target states
having either $2p_{3/2}$ orbit or both $2p_{3/2}$ and $1f_{5/2}$ active
orbits outside the inert core, $^{56}$Ni.\\
The targets being investigated with $^{40}$Ca as inert core  are
$^{42,44,46,48}$Ca, $^{45}$Sc, $^{46,48,50}$Ti, $^{51}$V,
$^{50,52,53,54}$Cr, $^{55}$Mn and $^{54,56,58}$Fe nuclei while those
investigated with
$^{56}$Ni as inert core include $^{58,60,61,62,64}$Ni, $^{63,65}$Cu,
$^{64,66}$Zn, $^{69}$Ga and $^{70}$Ge isotopes.\\
The new feature characterizing the availability of two active shells in
the target state  to transfer of a particle (hole) opens up many
interesting possibilities in the exploitation of experimental data on
transfer reactions. The study of targets having $^{40}$Ca as inert core
furnishes us with the $1f^2_{7/2}$,
$2p^2_{3/2}$ and $1f_{7/2}-2p_{3/2}$ effective interaction matrix
elements. A similar study of transfer reactions on targets having
$^{56}$Ni as inert core provides us with information regarding
$2p^2_{3/2}$, $1f^2_{5/2}$ and
$2p_{3/2}-1f_{5/2}$ effective  interaction.\\
The theoretical apparatus and the calculational procedure have been
discussed earlier in sufficient detail,\cite{r4,r5,r6} but for the sake of completeness
and convenience of the reader, we briefly reproduce the important
equations and other relevant features of our approach in the following
section.
\section {Sum Rule Equations and Method of Calculation}
The isospin centroids, $E^{\pm}_{T}$ (superscripts +,$-$ indicate
stripping and pick-up  cases respectively) of residual nuclear states
having isospin T, obtained via single particle stripping and
pick-up reactions on a target state with isospin $T_0$, are
given\cite{r4,r5} by\\
\noindent
$
E^{+}_{T_>}~-~E^+ (riz) $
$$=\frac
{\sum_k\{<H^{00}_{ik}>_{Tar}+<H^{01}_{ik}>_{Tar}+(N_i-\delta_{ik})q^{+}_{T_>}(
k)
\overline{W^{T=1}_{ik}}+(N_i+\delta_{ik})r^{+}_{T_>}(k)\overline
{W^{T=0}_{ik}}\}}{<\rho_i~neutron~ holes>_{Tar}}, \eqno{(1)}$$
\noindent
$
E^{+}_{T_<}~-~E^{+}(riz) $
$$= ~\frac{\sum_k { \{ <H^{00}_{ik}>_{Tar}-(\frac
{T_0+1}{T_0})<H^{01}_{ik}>_{Tar}+(N_i-\delta_{ik})q^{+}_{T_<}(k)\overline
{W^{T=1}_{ik}}+(N_i+\delta_{ik})r^{+}_{T_<}(k)\overline
{W^{T=0}_{ik}}\} } }
{<\rho_i~proton~holes>_{Tar}+\frac{1}{2T_0}\{<\rho_i~proton~holes
>_{Tar}-<\rho_i ~neutron~holes>_{Tar}\}}, \eqno{(2)}$$
$$ E^{-}_{T_>}~-~E^{-}(riz)
=\frac{\sum_k{\{<H^{00}_{ik}>_{Tar}-<H^{01}_{ik}>_{Tar}\}}}{<\rho_i~protons>_{
Tar}},~~~~~~~~~~~~~~~~~~~~~~~~~~~~~~~~~~~~~~~~~~~~~~~ ~~~  \eqno{(3)}$$
 and\\
\noindent
$ E^{-}_{T_<}~-~E^{-}(riz) $
$$= \frac{\sum_k {\{<H^{00}_{ik}>_{Tar}+(\frac
{T_0+1}{T_0})<H^{01}_{ik}>_{Tar}\}}}{<\rho_i~neutrons>_{Tar}+\frac
{1}{2T_0}\{<\rho_i~neutrons>_{Tar}-<\rho_i~protons>_{Tar}\}}
\eqno{(4)} $$
In these equations, $T_>~\equiv ~T_0+\frac{1}{2}$ ; $T_<~\equiv
 ~T_0-\frac{1}{2}$
 ; the summation index $k$ runs over all the active orbits in the
 target state, while $i$ refers to the $\rho_i(\equiv
 j\frac{1}{2})$ orbit
into (from) which the nucleon transfer occurs.  Further
$$ N_i~=~2j_i +1~~ ; \eqno{(5)}$$
$$q^{+}_{T}(k)~=~ \frac {3}{4}n_{k} + \frac
{f(T)T_{0k}}{2T_0}
 ~~ ; \eqno{(6)}$$
 $$ r^{+}_{T}(k)~=~\frac {1}{4}n_{k}- \frac
 {f(T)T_{0k}}{2T_0}~~;\eqno{(7)}$$
$n_k$ = number of nucleons in the $k$th active orbit in the
target state;
$$ f(T)~=~ T(T+1)-\frac {3}{4}-T_0(T_0+1) \equiv
\left\{ \begin{array}{ll}
T_0~for~T_> & \\
-(T_0+1)~for~T_<~~~; &
\end{array} \right.\eqno{(8)} $$
$T_{0k}$ = partial contribution of nucleons in the $k$th active
orbit towards the target state isospin;\\
$E^{\pm}(riz)~=~ E_0 \pm \epsilon_i$, with $E_0$ being the target
state energy and $\epsilon_i$, the single particle energy of
transferred nucleon with respect to the chosen inert core.\\
$\overline{W^{T=1}_{ik}}$ and $\overline {W^{T=0}_{ik}}$ in
equations (1) and (2) are $(2J+1)$-weighted averages of
two-body effective interaction matrix elements, $W^{JT}_{ikik}$, in isotriplet 
and isosinglet states,
respectively, of one nucleon in the $i$th orbit and another in
the $k$th orbit.
$$ <H^{00}_{ik}>_{Tar}=-\frac{1}{2} (1 +
\delta_{ik})E^{(2)}_{Tar}(i-k)\eqno{(9)} $$
where $E^{(2)}_{Tar}(i-k) $ is the total two-body interaction
energy of active nucleons in the $i$th orbit with those in the
$k$th orbit in the target state. $<H^{01}_{ik}>_{Tar}$ is the
isovector two-body correlation  term given by\\
\noindent
$ <H^{01}_{ik}>_{Tar}  $
$$ = \frac{1}{2} \sum_{\gamma}<Target
state\mid (2\gamma +1)^{1/2}W^{\gamma}_{ikik}[\{(A^{\rho_k }\times
A^{\rho_i})^{\gamma} \times B^{\rho_k}\}^{\rho_i} \times
B^{\rho_i}]^{01}\mid Target state>\eqno{(10)} $$
where the symbols $A^{\rho}$, $B^{\rho}$ etc. have their usual
meanings\cite{r4,r5}.  This term has, so far, defied an analytical
evaluation.  But as can be seen from equations (1) through (4),
this term can be eliminated by suitably combining any two of
these.\\
The isospin centroids, $E^{\pm}_T$, are calculated from experimental
data on excitation energies and spectroscopic strengths. While selecting
experimental data, we take into consideration  the fact that the total
strength for particle  transfer to/from a particular orbit compares
favourably with the non-energy weighted sum rules\cite{r7}.\\
Consistent with our basic assumptions and approach, given
earlier\cite{r5,r6}, we
like to mention that configuration mixing is not permissible in the
target states. As stated in our earlier works\cite{r4,r5,r6}, the
quantities, $n_k$,
$T_{0k}$ etc. and the denominators on right hand sides of equations (1)
to (4),
are  known from the chosen pure configuration for the target state
while $E^{\pm}(riz)$, $E^{(2)}_{Tar}$ etc. can be calculated with the
help of binding energy data. Then after eliminating the term
$<H^{01}_{ik}>_{Tar}$ as mentioned above, we  obtain, for each target
state, a linear equation involving the average interaction parameters,
$\overline {W}^T_{ik}$ as variables.
\section {Results and discussion}
We consider firstly the situation where a particle (hole) is transferred
to various target states with $^{40}$Ca as inert core. When the
transfer  of a particle (hole) takes place in the $1f_{7/2}$ orbit and
the target states are limited to $(1f_{7/2})^n$ configuration, it
enables us to extract $1f^2_{7/2}$ interaction matrix elements. This work
has been done earlier\cite{r5}. However, when the transfer is extended to
target states having two active orbits, i.e., $1f_{7/2}$ and $2p_{3/2}$,
it furnishes us with $1f^2_{7/2}$, $1f_{7/2}-2p_{3/2}$ and $2p^2_{3/2}$
interaction parameters.\\
When the inert core is shifted to $^{56}$Ni and the transfer  of a
particle (hole) takes place in the $2p_{3/2}$ orbit, either with
$2p_{3/2}$ active orbit alone or with $2p_{3/2}$ and $1f_{5/2}$ both as
active orbits, we are able to extract $2p^2_{3/2}$ and $2p_{3/2}-1f_{5/2}$
interaction parameters. When the transfer occurs in the $1f_{5/2}$ orbit
with appropriate target states, we can extract  $2p_{3/2}-1f_{5/2}$ as
well as $1f^2_{5/2}$ interaction matrix elements. It may, however, be
added that if the target states are  limited to $(2p_{3/2})^n$
configurations, then $2p_{3/2}$ transfer provides us with only the
$2p^2_{3/2}$ interaction parameters. \\
Thus, it is reasonably clear from the discussion above, that each
category of interaction  parameters can, in principle, be extracted  in
more than one ways.\\
The transfer
reactions\cite{r8,r9,r10,r11,r12,r13,r14,r15,r16,r17,r18,r19,r20,r21,r22,r23,r24,r25,r26,r27,r28,r29,r30,r31,r32,r33,r34,r35,r36,r37,r38,r39,r40,r41,r42,r43,r44,r45,r46,r47,r48,r49,r50,r51,r52,r53,r54,r55,r56,r57,r58,r59,r60,r61,r62,r63}
exploited in the present work with $^{40}$Ca and
$^{56}$Ni as inert cores are listed in tables I and II respectively.
Table III lists the values of effective interaction parameters,
$\overline {W}^{T=0}_{1f_{7/2}1f_{7/2}}$, $\overline
{W}^{T=1}_{1f_{7/2}1f_{7/2}}$, $\overline {W}^{T=0}_{1f_{7/2}2p_{3/2}}$,
$\overline {W}^{T=1}_{1f_{7/2}2p_{3/2}}$, $\overline
{W}^{T=0}_{2p_{3/2}2p_{3/2}}$ and $\overline
{W}^{T=1}_{2p_{3/2}2p_{3/2}}$ obtained from the calculations done with
$^{40}$Ca as inert core. Table IV lists the values of the effective
interaction parameters,$\overline {W}^{T=0}_{2p_{3/2}2p_{3/2}}$, $\overline
{W}^{T=1}_{2p_{3/2}2p_{3/2}}$, $\overline {W}^{T=0}_{2p_{3/2}1f_{5/2}}$,
$\overline {W}^{T=1}_{2p_{3/2}1f_{5/2}}$, $\overline
{W}^{T=0}_{1f_{5/2}1f_{5/2}}$ and $\overline
{W}^{T=1}_{1f_{5/2}1f_{5/2}}$ from calculations involving $^{56}$Ni as
inert core.\\
The extraction of the interaction matrix elements in all such single
particle transfer reaction problems and in particular, the present one,
suffers from the disadvantage that the experimental spectra and the
corresponding transfer reaction strengths are seldom available in full,
that is, in the spectra of the final nuclei, many energy states with
their corresponding strengths are not
seen by the experimentalists. This
limitation is further compounded by the fact that the shell model with
its pure configurational assumptions is not always strictly valid for
various nuclei. In spite of all these limitations, we have formulated our
equations\cite{r4,r5} under the assumption of pure configuration and then applied
these to the single particle transfer reaction data available to us.\\
While dealing with the cases of $1f_{7/2}$ and $2p_{3/2}$ transfer on
target states with $^{40}$Ca chosen as the inert core, we find in
literature, transfer reaction data for only five targets having both
$1f_{7/2}$ and $2p_{3/2}$ as active orbits. From these we are able to
set up only five  equations involving transfer of particle (hole) to a
particular orbit. Each of these equations contains four interaction
parameters  and from the algebraic point of view, five equations are not
enough to give us a good least-squares-fit for four parameters. We have,
therefore, for the case of $1f_{7/2}$ as well as $2p_{3/2}$ transfer,
combined the equations resulting from these five multishell target
states with other equations obtained from $(1f_{7/2})^n$
configuration target states; these we call as Calc. I and Calc. II,
respectively.\\
To increase the number of equations, for least-squares-fit purposes, to a
still larger number, we have done a fit for all the six interaction
parameters, $\overline {W}^{T=0}_{1f_{7/2}1f_{7/2}}$, $\overline
{W}^{T=1}_{1f_{7/2}1f_{7/2}}$, $\overline {W}^{T=0}_{1f_{7/2}2p_{3/2}}$,
$\overline {W}^{T=1}_{1f_{7/2}2p_{3/2}}$, $\overline
{W}^{T=0}_{2p_{3/2}2p_{3/2}}$ and $\overline
{W}^{T=1}_{2p_{3/2}2p_{3/2}}$, combining all the equations obtained from
$1f_{7/2}$ and $2p_{3/2}$ transfers; this we name as Calc. III. A
similar situation exists for
nuclei treated with $^{56}$Ni as inert core and here also each of the
cases of $2p_{3/2}$ and $1f_{5/2}$ transfer  has   been treated in a
similar manner as given above.\\
All the results are listed in tables III and IV alongwith those of other
authors\cite{r64,r65,r66,r67,r68,r69}. The results of present  work in the three categories under
Calc. I, II \& III, as mentioned above in the discussion, show minor
variations; the magnitude of variations lies well within the various
uncertainties caused by the discrepancies in the spectroscopic strengths
and other experimental measurements.
\acknowledgements
We would like to thank the CSIR (India) for the financial support to this
work.

\vfill \eject
\begin{center}
Table I. Single particle transfer reaction data used for evaluation of
$1f^2_{7/2}$ and $1f_{7/2}-2p_{3/2}$ average interaction matrix elements
with $^{40}$Ca as inert core.
\begin{tabular}{l c c c c c c} \hline \hline
Target & Stripping & \multicolumn{2}{c}{Centroid}
& Pick-up & \multicolumn{2}{c}{Centroid}\\ \cline{3-4}
\cline{6-7}
& Reaction & Isospin & Value & Reaction & Isospin &
Value\\
& [Reference]& & (MeV) & [Reference] & & (MeV)\\ \hline
& &   \multicolumn{2}{c}{a) $1f_{7/2}$ transfer}&  & &\\
&&&&&&\\
$^{42}$Ca & $(d,p)^{8)}$ & $T_>$ &0.066 & $(d,t)^{9)}$& $T_<$&0.000\\
$^{42}$Ca & $(^{3}$He,$d)^{10)}$ & $T_<$ &1.079 &$(p,d)^{11)}$& $T_<$&
0.567\\
$^{44}$Ca & $(^{3}$He,$d)^{12)}$ & $T_<$ & 0.453 & $(p,d)^{11)}$ &$T_<$
&0.000\\
$^{45}$Sc & $(d,p)^{13)}$ & $T_>$ & 0.383 & $(d$, $^{3}$He)$^{14)}$ & $T_>$
& 1.086\\
$^{46}$Ca & $(d,p)^{15)}$ & $T_>$ & 0.000 &$(d,t)^{9)}$&$T_<$ &0.000\\
$^{46}$Ti & $(^{3}$He,$d)^{16)}$ & $T_<$ & 0.150 &$(p,d)^{17)}$ &$T_<$
&0.434\\
$^{46}$Ti & $(d,p)^{18)}$ & $T_>$ & 0.555 &$(p,d)^{17)}$ &$T_>$
&4.760\\
$^{48}$Ca & $(^{3}$He,$d)^{19)}$ & $T_<$ &0.000 &$(p,d)^{11)}$ & $T_<$ &
0.050\\
$^{48}$Ti & $(d,p)^{20)}$ & $T_>$ & 0.583 & $(d$,$^{3}$He)$^{21)}$&
$T_>$ & 0.000\\
$^{50}$Cr & $(d,p)^{22)}$ & $T_>$ &0.302 & $(t,\alpha)^{23)}$ & $T_>$ &
0.490\\
$^{51}$V & $(^{3}$He,$d)^{24)}$ & $T_<$ & 2.279 & $(p,d)^{25)}$ & $T_<$
& 1.769\\
$^{53}$Cr & $(^{3}$He,$d)^{26)}$ & $T_<$ & 0.256&
$(^{3}$He,$\alpha)^{27)}$ & $T_<$ & 4.095\\
$^{54}$Cr & $(^{3}$He,$d)^{16)}$ & $T_<$ & 0.126 &
$(^{3}$He,$\alpha)^{28)}$&$T_<$ & 1.891\\
$^{55}$Mn & $(\alpha,t)^{29)}$ & $T_<$ & 2.075 & $(d,t)^{30)}$ & $T_<$
& 1.950\\
$^{56}$Fe & $(^{3}$He,$d)^{31)}$ & $T_<$ & 1.244 & $(d,t)^{32)}$ & $T_<$
& 1.384\\
$^{58}$Fe & $(\alpha,t)^{33)}$ & $T_<$ & 0.000& $(p,d)^{25)}$ & $T_<$
&2.220\\
& & & & & & \\
& & \multicolumn{2}{c}{b) $2p_{3/2}$ transfer}&  & &\\
& & & & & & \\
$^{48}$Ca & $(^{3}$He,$d)^{19)}$ & $T_<$ & 3.536 & --- &---  &---\\
$^{46}$Ti & $(d,p)^{18)}$ & $T_>$ & 2.438 & --- &---&---\\
$^{48}$Ti & $(d,p)^{20)}$ & $T_>$ & 2.080 & --- & --- &---\\
$^{50}$Ti & $(d,p)^{34)}$ & $T_>$ & 0.389 & --- & --- &---\\
$^{50}$Ti & $(^{3}$He,$d)^{35)}$ & $T_<$ & 2.636 & --- &---&---\\
$^{50}$Cr & $(d,p)^{36)}$ & $T_>$ & 2.080 & --- & --- &---\\
$^{50}$Cr & $(^{3}$He,$d)^{16)}$ & $T_<$ & 2.387 & --- &---&---\\
$^{52}$Cr & $(^{3}$He,$d)^{37)}$ & $T_<$ & 2.316 & --- &---&---\\
$^{52}$Cr & $(d,p)^{38)}$ & $T_>$ & 0.773 & --- & --- &---\\
$^{54}$Fe & $(^3$He,$d)^{39)}$ & $T_>$ & 4.752 & --- & --- &---\\
$^{54}$Fe & $(^{3}$He,$d)^{39)}$ & $T_<$ & 2.453 & --- &---&---\\
$^{54}$Cr & $(^{3}$He,$d)^{16)}$ & $T_<$ & 2.523 &
$(^{3}$He,$\alpha)^{27)}$ &$T_<$ & 0.000\\
$^{54}$Cr & $(d,p)^{40)}$ & $T_>$ & 0.000 &$(^{3}$He,$\alpha)^{27)}$ &
$T_<$ & 0.000\\
$^{55}$Mn & $(d,p)^{41)}$ & $T_>$ & 0.170 & $(d,t)^{30)}$ &$T_<$&0.120\\
$^{56}$Fe & $(^{3}$He,$d)^{31)}$ & $T_<$ & 1.454 &$(p,d)^{42)}$ &$T_<$
&0.000\\
$^{56}$Fe & $(d,p)^{43)}$ & $T_>$ & 0.147 &$(p,d)^{42)}$  &$T_<$ &0.000\\
\hline
\end{tabular}
\end{center}
\vfill\eject
\begin{center}
Table II. Single particle transfer reaction data used for evaluation of
$2p^2_{3/2}$ and $1f_{7/2}-2p_{3/2}$ average interaction matrix elements
with $^{56}$Ni as inert core.
\begin{tabular}{l c c c c c c} \hline \hline
Target & Stripping & \multicolumn{2}{c}{Centroid}
& Pick-up & \multicolumn{2}{c}{Centroid}\\ \cline{3-4}
\cline{6-7}
& Reaction & Isospin & Value & Reaction & Isospin &
Value\\
& [Reference]& & (MeV) & [Reference] & & (MeV)\\ \hline
& &   \multicolumn{2}{c}{a) $2p_{3/2}$ transfer}&  & &\\
&&&&&&\\
$^{58}$Ni & $(^{3}$He, $d)^{44)}$ & $T_<$ &0.000 & $(p,d)^{45)}$& $T_<$&0.000\\
$^{58}$Ni & $(d,p)^{46)}$ & $T_>$ &0.000 &$(p,d)^{45)}$& $T_<$&0.000\\
$^{60}$Ni & $(^{3}$He,$d)^{47)}$ & $T_<$ & 0.000 & $(p,d)^{48)}$ &$T_<$
&0.000\\
$^{62}$Ni & $(^{3}$He, $d)^{49)}$ & $T_<$ & 0.000 & $(p,d)^{50)}$ &
$T_<$& 0.166\\
$^{64}$Ni & $(^{3}$He,$d)^{49)}$ & $T_<$ & 0.000 &$(d,t)^{30)}$&$T_<$
&0.250\\
$^{63}$Cu & $(^{3}$He,$d)^{51)}$ & $T_<$ & 1.810 &$(d,^{3}$He$)^{14)}$
&$T_>$&0.492\\
$^{65}$Cu & $(^{3}$He, $d)^{51)}$ & $T_<$ & 1.646 &$(d,^{3}$He$)^{14)}$
&$T_>$&0.354\\
$^{64}$Zn & $(^{3}$He,$d)^{52)}$ & $T_<$ &0.251 &$(d,^{3}$He$)^{53)}$ &
$T_>$&0.060\\
$^{66}$Zn & $(^{3}$He,$d)^{54)}$ & $T_<$ & 0.263 &
$(^{3}$He,$\alpha)^{54)}$& $T_<$ & 0.171\\
& & & & & & \\
& & \multicolumn{2}{c}{b) $1f_{5/2}$ transfer}&  & &\\
& & & & & & \\
$^{61}$Ni & $(d,p)^{55)}$ & $T_>$ & 3.141 &$(p,d)^{50)}$ &$T_<$&2.517\\
$^{62}$Ni & $(^{3}$He,$d)^{49)}$ & $T_<$ & 1.437 &$(p,d)^{50)}$& $T_<$&0.227\\
$^{62}$Ni & $(d,p)^{56)}$ & $T_>$ & 0.087 &$(p,d)^{50)}$& $T_<$&0.227\\
$^{63}$Cu & $(^{3}$He,$d)^{51)}$ & $T_<$ & 2.955 & $(d,t)^{57)}$&$T_<$
&0.530\\
$^{64}$Ni & $(^{3}$He,$d)^{49)}$ & $T_<$ & 1.684 & $(d,t)^{30)}$
&$T_<$&0.087\\
$^{64}$Ni & $(d,p)^{58)}$ & $T_>$ & 0.000 & $(d,t)^{30)}$
&$T_<$&0.087\\
$^{64}$Zn & $(^{3}$He,$d)^{52)}$ & $T_<$ & 0.626 &$(p,d)^{59)}$ &$T_<$&0.193\\
$^{66}$Zn & $(^{3}$He,$d)^{54)}$ & $T_<$ & 0.468
&$(^{3}$He,$\alpha)^{54)}$ &$T_<$&0.026\\
$^{69}$Ga & $(^{3}$He, $d)^{60)}$ & $T_<$ & 2.834 &$(d,t)^{61)}$ &
$T_<$ &0.415\\
$^{70}$Ge & $(^{3}$He,$d)^{62)}$ & $T_<$ & 0.000 & $(d,t)^{63)}$ &$T_<$
&0.000\\
\hline
\end{tabular}
\end{center}
\vfill\eject
\begin{center}
Table III. Average interaction parameters with $^{40}$Ca as inert core.
All values in MeV.
\begin{tabular}{l c c c c c c } \hline \hline
& & & & & &       \\
& $\overline{W}^{T=0}_{1f_{7/2}1f_{7/2}}$ &
$\overline{W}^{T=1}_{1f_{7/2}1f_{7/2}}$&
$\overline{W}^{T=0}_{1f_{7/2}2p_{3/2}}$&
$\overline{W}^{T=1}_{1f_{7/2}2p_{3/2}}$&
$\overline{W}^{T=0}_{2p_{3/2}2p_{3/2}}$&
$\overline{W}^{T=1}_{2p_{3/2}2p_{3/2}}$\\ \hline
Present work Calc. I$^{a}$& $-$1.678  & $-$0.208  &$-$1.979 & 0.275 &---
&--- \\
~~~~~~~~~~~~~~~~~~Calc. II$^{b}$ &--- &---& $-$1.754 &0.278& $-$1.285 &
$-$0.526\\
~~~~~~~~~~~~~~~~~Calc. III$^{c}$ &$-$1.724 &$-$0.204& $-$1.846 &0.305&
$-$1.165 &$-$0.560\\
Previous work Calc. I$^{d}$ &$-$1.714 &$-$0.215& $-$1.738 & 0.267&
$-$1.503 &$-$0.255\\
~~~~~~~~~~~~~~~~~~Calc. II$^{e}$ &$-$1.662 &$-$0.212&---&--- &--- &---\\
Richter et al. FPD6$^{f}$ & $-$1.462 &$-$0.208 &$-$1.201 & 0.155
&$-$1.455 &0.011\\
Richter et al.FPMI3$^{f}$ & $-$1.095 &$-$0.191 &$-$1.077 & 0.145&
$-$1.522 &$-$.523\\
Kuo and Brown$^{g}$ & $-$1.154 &$-$0.128 &$-$0.968 & $-$0.104 &$-$1.474 &
$-$0.518\\

Schiffer \& True  Set I$^{h}$ & $-$1.594 &--- &--- &--- &--- &---\\
~~~~~~~~~~~~~~~~~~~~Set II$^{h}$& $-$1.739 &--- &--- &--- &--- &---\\

Lips \& McEllistren $^{i}$&---&$-$0.240 &--- &$-$0.501 &--- &---\\
Federman \& Pittel$^{j}$&---&$-$0.228 &--- &$-$0.151 &--- &---\\ \hline
\end{tabular}
\end{center}
\vskip .5cm
\noindent
$^{a}$ From $1f_{7/2}$ transfer on target states having $1f_{7/2}$ alone
as well as  $1f_{7/2}$ and $2p_{3/2}$ both as active  orbits.\\
$^{b}$ From $2p_{3/2}$ transfer on target states having $1f_{7/2}$ alone
as well as  $1f_{7/2}$
and $2p_{3/2}$ both as active orbits.\\
$^{c}$ From a least-squares fit on the combined data of a) \& b).\\
$^{d}$ From transfer reactions on target states having only neutron
occupancy in the active orbit involved in transfer\cite{r3}.\\
$^{e}$ From $1f_{7/2}$ transfer  on target states having only
$1f_{7/2}$ as the active orbit\cite{r5}.\\
$^{f}$ ref 64;  $^{g}$ ref 65; $^{h}$ ref 66; $^{i}$ ref 67; $^{j}$ ref
68.
\vfill \eject
\begin{center}
Table IV. Interaction parameters with $^{56}$Ni as inert core.
All values in MeV.
\begin{tabular}{l c c c c c c } \hline \hline
& & & & & &       \\
& $\overline{W}^{T=0}_{2p_{3/2}1p_{3/2}}$ &
$\overline{W}^{T=1}_{2p_{3/2}2p_{3/2}}$&
$\overline{W}^{T=0}_{2p_{3/2}1f_{5/2}}$&
$\overline{W}^{T=1}_{2p_{3/2}1f_{5/2}}$&
$\overline{W}^{T=0}_{1f_{5/2}1f_{5/2}}$&
$\overline{W}^{T=1}_{1f_{5/2}1f_{5/2}}$\\ \hline
Present work Calc. I$^{a}$& $-$1.460  & $-$0.226  &$-$1.594 & 0.449 &---  &--- \\
~~~~~~~~~~~~~~~~~~Calc. II$^{b}$ &--- &---& $-$1.811 & 0.491& $-$1.272 &
$-$0.300\\
~~~~~~~~~~~~~~~~~~Calc. III$^{c}$ &$-$1.424 &$-$0.204& $-$1.724 & 0.499& $-$1.341 &
$-$0.309\\
Previous work calc I$^{d}$ &$-$1.503 &$-$0.255& $-$2.163 & 0.432&
$-$1.235 &$-$0.179\\
Richter et al. FPD6$^{e}$ & $-$1.455 &0.105 &$-$1.223 & 0.093
&$-$1.343 &$-$0.087\\
Richter et al.FPMI3$^{e}$ & $-$1.522 &$-$0.523 &$-$0.941 & $-$0.017&
$-$0.952 &0.045\\
Kuo and Brown$^{f}$ & $-$1.474 &$-$0.518 &$-$0.970 & $-$0.007 &$-$0.951 &
0.049\\
Glaudemans et al$^{g}$ & --- &0.111 &--- & 0.200 &--- &
0.147\\  \hline
\end{tabular}
\end{center}
\vskip .5cm
\noindent
$^{a}$ From $2p_{3/2}$ transfer on target states having $2p_{3/2}$ alone
as well as $2p_{3/2}$ and $1f_{5/2}$ both as active  orbits.\\
$^{b}$ From $1f_{5/2}$ transfer on target states having $2p_{3/2}$ alone
as well as  $2p_{3/2}$ and $1f_{5/2}$ both as active orbits.\\
$^{c}$ From  least-squares fit on combined data of a) \& b).\\
$^{d}$ From transfer reactions on target states having only neutron
occupancy in the active  orbit involved in transfer\cite{r3}.\\

$^{e}$ ref 64; $^{f}$ ref 65; $^{g)}$ ref 69.

\begin{thebibliography}{90}
\baselineskip 24pt
\bibitem{r1} R. K. Bansal and J. B. French, Phys. Lett. {\bf
11} (1964), 145.
\bibitem{r2} R. K. Bansal, Phys. Lett. {\bf 40B} (1972), 189.
\bibitem{r3} R. K. Bansal and Ashwani Kumar, Pramana - J. Phys.
 (India) {\bf 9} (1977), 273.
\bibitem{r4} R. K. Bansal and Ashwani Kumar, Pramana - J.
Phys. (India) {\bf 32} (1989), 341.
\bibitem{r5} R. K. Bansal, H. Sharda and Ashwani Kumar, Phys.
Lett. {\bf 386B} (1996), 17.
\bibitem{r6} H. Sharda, R. K. Bansal and Ashwani Kumar, Prog. Theor.
Phys. {\bf 100} (1998) 737.
\bibitem{r7} J. B. French and M. H. Macfarlane, Nucl. Phys.{\bf 26}
(1961), 168.
\bibitem{r8} G. Brown, A. Denning and J. G. B. Haigh , Nucl.
Phys.{\bf A224} (1974), 267.
\bibitem{r9} J. L. Yntema, Phys. Rev. {\bf 186}
(1969), 1144.
\bibitem{r10} J. Bommer, K. Grabisch, H. Kluge, G.
Roschert, H. Fuchs, U. Lynen and Kamal K. Seth, Nucl. Phys. {\bf A160}
(1971), 577.
\bibitem{r11} P. Martin, M. Buenerd, Y. Dupont and M.
Chabre, Nucl. Phys. {\bf A185}
(1972), 465.
\bibitem{r12} J. J. Schwartz and W. Parker Alford, Phys.
Rev.{\bf 149} (1966), 820.
\bibitem{r13} J. Rapaport, A. Sperduto and W. W. Buechner,
Phys. Rev. {\bf 151} (1966), 939.
\bibitem{r14} G. Mairle, G. Th. Kaschl, H. Link, H. Mackh,
U. Schmidt-Rohr and G. J. Wagner, Nucl. Phys. {\bf A134} (1969), 180.
\bibitem{r15} J. H. Bjerregaard, Ole Hansen and  G.
Sidenius, Phys. Rev. {\bf 138} (1965), B1097.
\bibitem{r16} B. Cujec and I. M. Szoghy, Phys. Rev. {\bf
179} (1969), 1060.
\bibitem{r17} P. J. Plauger and E. Kashy, Nucl. Phys. {\bf
A152} (1970), 609.
\bibitem{r18} M. S. Chowdhury and H. M. Sen Gupta, Nucl.
Phys. {\bf A229} (1974), 484.
\bibitem{r19} J. R. Erskine, A. Marinov and J. P.
Schiffer, Phys. Rev. {\bf 142} (1966), 633.
\bibitem{r20} A. E. Ball, G. Brown, A. Denning and R. N.
Glover, Nucl. Phys. {\bf A183} (1972), 472.
\bibitem{r21} H. Ohnuma, Phys. Rev. {\bf C3} (1971), 1192.
\bibitem{r22} A. E. Macgregor and G. Brown, Nucl. Phys. {\bf A190}
(1972), 548.
\bibitem{r23} D. Bachner, R. Santo, H. H. Duhm, R. Bock and S.
Hinds, Nucl.  Phys. {\bf A106} (1968), 577.
\bibitem{r24} M. A. Basher, H. R. Siddique, A. Husain, A. K. Basak
and H. M. Sen Gupta, Phys. Rev. {\bf C45} (1992), 1575.
\bibitem{r25} J. C. Legg and E. Rost, Phys. Rev. {\bf 134}
(1964),B752.
\bibitem{r26} L. L. Lynn, W. E. Dorenbusch, T. A. Belote and J.
Rapaport, Nucl. Phys. {\bf A135} (1969), 97.
\bibitem{r27} P. David H. H. Duhm, R. Bock and R. Stock, Nucl. Phys.
{\bf A128} (1969), 47.
\bibitem{r28} S. Fortier, E. Hourani, M. N. Rao and S. Gales, Nucl.
Phys. {\bf A311} (1978), 324.
\bibitem{r29} Masaru Matoba, Nucl. Phys. {\bf A118} (1968), 207.
\bibitem{r30} J. A. Cameron, E. Habib, A. A. Pilt, R. Schubank and
V. Janzen, Nucl. Phys. {\bf A365} (1981), 113.
\bibitem{r31} G. Hardie, T. H. Braid, L. Meyer-Schutzmeister and J.
W. Smith, Phys. Rev. {\bf C5} (1972), 1600.
\bibitem{r32} W. P. Alford, J. A. Cameron, E. Habib, Phys. Rev. {\bf
C44} (1991), 319.
\bibitem{r33} D. D. Armstrong, A. G. Blair, H. C. Thomas, Phys. Rev.
{\bf 155} (1967), 1254.
\bibitem{r34} D. C. Kocher and W. Haeberli, Nucl. Phys. {\bf A196}
(1972), 225.
\bibitem{r35} P. N. Maheshwari, U. C. Gupta, C. St-Pierre, Can. J.
Phys. {\bf 49} (1971), 1053.
\bibitem{r36} G. Delic and B. A. Robson, Nucl. Phys. {\bf A134}
(1969), 470.
\bibitem{r37} J. E. Park, W. W. Daehnick and M. J. Spisak, Phys. Rev.
{\bf C19} (1979), 42.
\bibitem{r38} G. Brown, A. Denning, A. E. Macgregor, Nucl. Phys.
{\bf A153} (1970), 145.
\bibitem{r39} O. Karban, A. K. Basak, F. Entezami, P. M. Lewis and
S. Roman, Nucl. Phys. {\bf A369} (1981), 38.
\bibitem{r40} T. Taylor and J. A. Cameron, Nucl. Phys. {\bf A337}
(1980), 389.
\bibitem{r41} J. R. Comfort, Phys. Rev. {\bf 177} (1969), 1573.
\bibitem{r42} K. Hosono, M. Kondo, T. Saito, N. Matsuoka, S.
Nagamachi, T. Noro, H. Shimizu, S. Kato, K. Okada, K. Ogino and Y.
Kadota, Nucl. Phys. {\bf A343} (1980), 234.
\bibitem{r43} James A. Thomson, Nucl. Phys. {\bf A227} (1974),
485.
\bibitem{r44} P. K. Bindal, D. H. Youngblood and R. L. Kozub, Phys.
Rev. {\bf C14} (1976), 521.
\bibitem{r45} F. M. Edwards, J. J. Kraushaar and B. W. Ridley, Nucl.
Phys. {\bf A199} (1973), 463.
\bibitem{r46} M. S. Chowdhury and H. M. Sen Gupta, Nucl. Phys. {\bf
A205} (1973), 454.
\bibitem{r47} J. E. Kim and W. W. Daehnick, Phys. Rev. {\bf C24}
(1981), 1461.
\bibitem{r48} H. Nann, D. W. Miller, W. W. Jacobs, D. W. Devins, W.
P. Jones and A. G. M. Van Hees, Phys. Rev. {\bf C28} (1983), 642.
\bibitem{r49} R. M. Britton and D. L. Watson, Nucl. Phys. {\bf A272}
(1976), 91.
\bibitem{r50} D. H. Koang, W. S. Chien and H. Rossner, Phys. Rev. {\bf
C13} (1976), 1470.
\bibitem{r51}  J. L. C. Ford, Jr., K. L. Warsh, R. L. Robinson and C.
D. Moak, Nucl. Phys. {\bf A103} (1967), 525.
\bibitem{r52} B. Zeidman, R. H. Siemssen, G. C. Morrison and L. L.
Lee., Jr., Phys. Rev. {\bf C9} (1974), 409.
\bibitem{r53} Ole Hansen, M. N. Harakeh, J. V. Maher, L. W. Put and
J. C. Vermeulen, Nucl. Phys. {\bf A313} (1979), 95.
\bibitem{r54} M. G. Betigeri, P. David, J. Debrus, H. Mommsen and A.
Riccato, Nucl. Phys. {\bf A171} (1971), 401.
\bibitem{r55} O. Karban, A. K. Basak, F. Entezami and S. Roman,
Nucl. Phys.{\bf A366} (1981), 68.
\bibitem{r56} G. A. Huttlin, S. Sen, W. A. Yoh and A. A. Rollefson,
Nucl. Phys.{\bf A227} (1974), 389.
\bibitem{r57} W. W. Daehnick, Y. S. Park and D.L. Dittmer, Phys. Rev.
{\bf C8} (1973), 1394.
\bibitem{r58} T. R. Anfinsen, K. Bjorndal, A. Graue, J. R. Lien, G.
E. Sanvik, L. O. Tveita, K. Ytterstad and E. R. Cosman, Nucl. Phys. {\bf
A157} (1970), 561.
\bibitem{r59} P. A. S. Metford, T. Taylor and J. A. Cameron, Nucl.
Phys. {\bf A308} (1978), 210.
\bibitem{r60} D. Ardouin, R. Tamisier, M. Verges, G. Rotbard, J.
Kalifa, G. Berrier and B. Grammaticos, Phys. Rev. {\bf C12} (1975), 1745.
\bibitem{r61} W. W. Daehnick, S. Shastry, M. J. Spisak and D. Gur,
Phys. Rev. {\bf C15} (1977), 1264.
\bibitem{r62} R. R. Betts, S. Mordechai, D. J. Pullen, B. Rosner and  W.
Scholz, Nucl. Phys. {\bf A230} (1974), 235.
\bibitem{r63} J. A. Bieszk, Luis Montestruque and S. E. Darden,
Phys. Rev. {\bf C16} (1977), 1333.
\bibitem{r64} W. A. Richter, M. G. Van Der Merwe, R. E. Julies and B.
A. Brown, Nucl. Phys. {\bf A523} (1991), 325.
\bibitem{r65} T. T. S. Kuo and G. E. Brown, Nucl. Phys. {\bf A114}
(1968), 241.
\bibitem{r66} J. P. Schiffer and W. W. True, Rev. Mod. Phys. {\bf 48}
(1976), 191.
\bibitem{r67} K. Lips and M. T. McEllistren, Phys. Rev. {\bf C1}
(1970), 1009.
\bibitem{r68} P. Federman and S. Pittel, Nucl. Phys. {\bf A155}
(1970), 161.
\bibitem{r69} P. W. M. Glaudemans, M. J. A. De Voigt and E. F. M.
Steffens, Nucl. Phys. {\bf A198} (1972), 609.
\end{thebibliography}
\end{document}